\documentclass{iau}

\usepackage{amsmath}
\usepackage{graphicx}
\usepackage{multirow}
\usepackage[capitalise]{cleveref}

\begin{document}

\lefttitle{Timo Kravtsov}
\righttitle{Environments of type Ibn SNe}

\jnlPage{1}{7}
\jnlDoiYr{2026}
\doival{10.1017/xxxxx}

\aopheadtitle{Proceedings IAU Symposium}
\volno{406}
\editors{H. Kuncarayakti, T. Kangas \& M. Bersten, eds.}

\title{The environments of type Ibn supernovae: evidence for a strong Wolf-Rayet star component}

\author{T.~Kravtsov$^{1,2}$, H.~Kuncarayakti$^{1,2}$, J.~P.~Anderson$^{3}$}
\affiliation{$^1$ Tuorla Observatory, Department of Physics and Astronomy, University of Turku, 20014 Turku, Finland}
\affiliation{$^2$ Finnish Centre for Astronomy with ESO (FINCA), University of Turku, Vesilinnantie 5, Quantum 20014 Turku, Finland}
\affiliation{$^3$ European Southern Observatory (ESO), Alonso de Córdova 3107, Casilla 19, Santiago 19001, Chile}

\begin{abstract}
Characterisation of progenitor stars of supernovae (SNe) poses several challenges. Direct observations of progenitors are based on serendipitious archival observations and are only possible for nearby events. Indirect methods include inferring nickel mass from the light curves or oxygen masses from the SN's nebular phase, both of which are tools to constrain the progenitor's initial mass. Alternatively, environments of SNe have been used as a proxy for the progenitor's age and metallicity. The progenitor channel of type Ibn SNe is a hotly debated topic. The mass-loss and the formation of He-rich circumstellar medium requires strong envelope stripping of the progenitor star. For core-collapse supernovae (CCSNe), this requires strong stellar wind mass losses, i.e. a Wolf-Rayet (WR) star, or binary interaction with a companion star. Here, we analyse a sample of host galaxies of type Ibn SNe that have been observed with MUSE. We use H$\alpha$ emission to infer the transient's association to star formation and derive the local metallicity with commonly used strong line diagnostic. We find that the sample of SNe are highly associated with star formation in the host galaxy, typically situated within one of the brightest HII regions in the host galaxy. The NCR distribution of H$\alpha$ pixels in SN Ibn environments strongly resembles that of WR stars, and shows a stronger preference to SF compared to other core-collapse SN types. Metallicity measurements show typically sub-solar metallicity in the SN environments.

\end{abstract}


\maketitle

\section{Introduction}
A central question in the study of supernovae (SNe) is the connection of SN classes to their respective progenitors. While some stars have been observed before their eventual cataclysmic end, this approach always requires some degree of luck and has been achieved in only a handful of cases. The most famous example is the blue supergiant Sk~-69~202 in the Large Magellanic Cloud that exploded as SN~1987A and was observed by surveys in the preceding decades \citep{1987A&A...177L...1W}. More recently, pre-explosion images were obtained for SN~2019yvr, which showed a yellow supergiant as the likely progenitor star \citep{2021MNRAS.504.2073K}. Pre-explosion imaging is also limited by current instrumentation as individual stars become harder to distinguish with distance.

The evolution of the explosion itself may glean us information about the star that produced it. $^{56}$Ni created during the SN explosion decays into $^{56}$Co, which produces much of the light curve of the SN. By estimating the amount of $^{56}$Ni produced in the explosion from the light curve shape, the mass of the progenitor may be constrained. During later phases the photospehere of the SN recedes and the SN enters its nebular phase. In this phase, information on the progenitor may be constrained with oxygen abundances derived from the oxygen emission lines \citep{2015A&A...579A..95K}.

Observationally the least time-critical method is the use of SN environments as an information source for the progenitor; here the SN itself is a contaminant for observations. Progenitors of core-collapse SNe (CCSNe) are massive and short-lived stars. The lifespan is not long enough for them to disperse from the original group of stars with which they were originally born. Therefore, by studying the immediate stellar environments of SNe information on the progenitor may be constrained. Stellar population age and gas-phase metallicity of the interstellar medium (ISM) directly hint at the lifespan and metallicity of the progenitor.

\section{Observations and Methods}
We observed 24 hosts of type Ibn SNe with the Multi Unit Spectroscopic Explorer (MUSE) instrument on the Very Large Telescope. MUSE is a medium-resolution integral field spectrograph with a resolving power $R\sim$3000 and a wavelength range of 4650-9300 \AA. Data included in the analysis are from European Southern Observatory (ESO) Programme ID 0111.B-0149. Each host galaxy observation consists of 4$\times$ 700 s observations. The data was reduced using the standard ESO pipeline EsoReflex. The programme ran as a poor-weather filler and the image quality in the dataset is typically $\gtrsim$1".

MUSE typically has a random $\sim$1" offset in its World Coordinate System (WCS) solution from the reduction pipeline. We calibrated each datacube's WCS with stars in the field of view, using reference stars from the Two Micron All-Sky Survey (2MASS) and Gaia Data Release 2. SN locations were then identified using the coordinates reported the Transient Name Server\footnote{https://www.wis-tns.org} and Open Supernova Catalog\footnote{sne.space}\citep{2017ApJ...835...64G}. After the identification we extracted a spectrum from the datacube with a circular aperture with the seeing of the observation as a the diameter. The flux of the extracted spectrum was then divided by the number of pixels in the aperture. This produces a spectrum that has the representative flux scale to the other spaxel but is still representative of the emission surrounding the SN. Additionally, it accounts for the uncertainties in the identification of the SN location and increases the signal-to-noise of the SN environment observation.

Our methods concentrate on the gas-phase emissions of the host galaxies and therefore the stellar continuum is removed. We use \texttt{STARLIGHT} to fit simple stellar population models from BPASS v2.2 to identify the continuum emission from stellar populations in each spaxel. The continuum is then subtracted and only the gas emission lines are left in the spaxels. We use the Balmer decrement to deredden the data by assuming Case B recombination and intrinsic H$\alpha$/H$\beta$ ratio of 2.86. Finally, we create strong line emission maps based on the dereddened emission-only datacubes.

We adopt H$\alpha$ brightness as an age indicator. H$\alpha$ emission in HII regions is produced from the ionising radiation of massive stars. Therefore, it is associated with recent star formation \citep{1998ARA&A..36..189K}. We use Normalised Cumulative Rank (NCR) to identify if the SNe are prefenterially in star forming regions of its host galaxy \citep{2008MNRAS.390.1527A, 2012MNRAS.424.1372A}. In short, we assign a value between 0 and 1 to pixels so that the pixel with lowest H$\alpha$ value is at zero and brightest pixel is at unity. In essence, each pixel receives a value that represents its H$\alpha$ brightness as a fraction of the brightest pixel. A corresponding NCR value is also calculated for the SN spectrum.

We measure metallicity of the SN parent HII region with a strong line diagnostic that uses [N\,II]$\lambda$6584/H$\alpha$ line ratio as a metallicity measurement tool \citep{2013A&A...559A.114M}. These two lines are typically the brightest optical light emission lines and maximises the number of pixels where the measurement may be carried out.

\section{Results}
\begin{figure}
    \centering
    \includegraphics[width=0.5\linewidth]{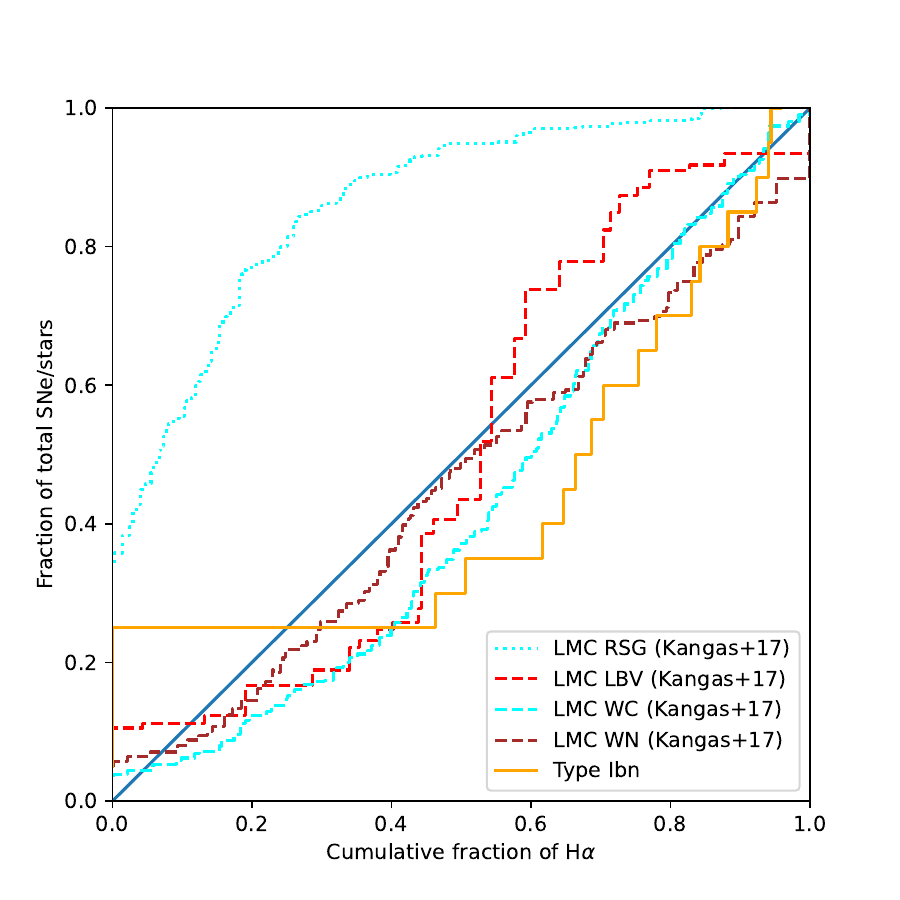}
    \caption{CDF of H$\alpha$ measurement compared to star classes from \cite{2017A&A...597A..92K}.}
    \label{fig:kangas}
\end{figure}

\begin{figure}
    \centering
    \includegraphics[width=0.5\linewidth]{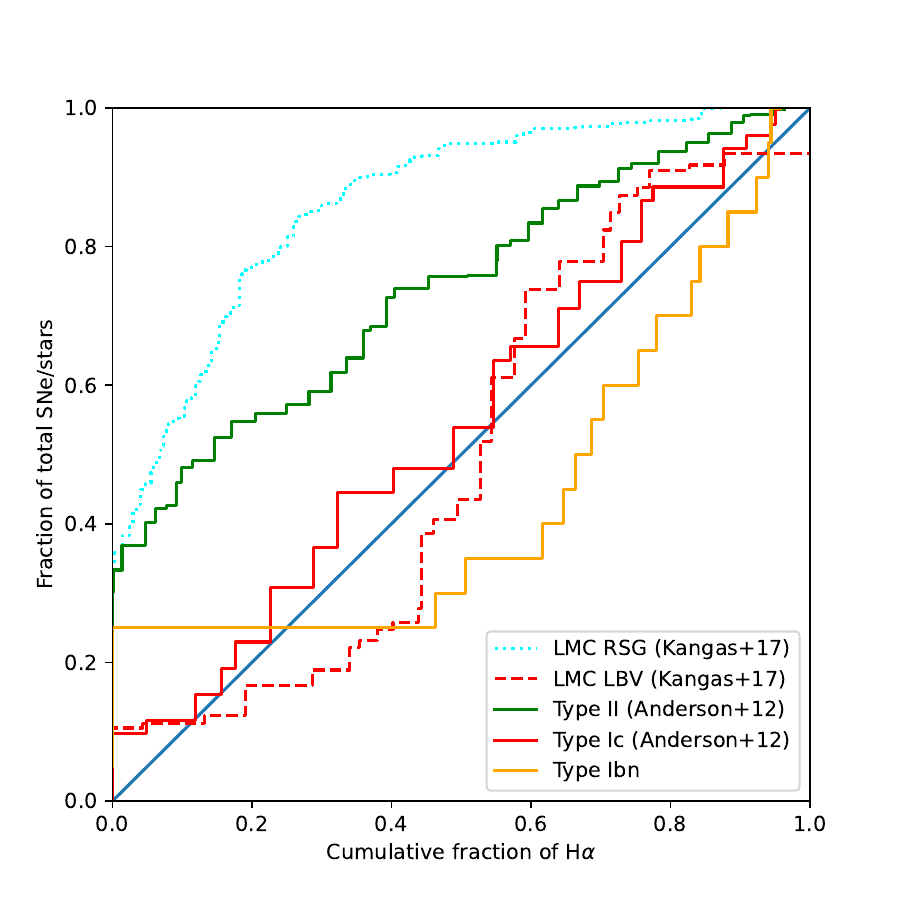}
    \caption{Same as \cref{fig:kangas}, but with addition of type II and type Ic CDFs from \cite{2012MNRAS.424.1372A}.}
    \label{fig:anderson}
\end{figure}

We find that type Ibn SNe are strongly associated with bright H$\alpha$ regions. Qualitatively they are typically in one of the brightest HII regions in the host galaxy. In \cref{fig:kangas}, we present our cumulative distribution function (CDF) compared to CDFs of a sample of stellar classes from \cite{2017A&A...597A..92K}. We find that type Ibn SNe most strongly correlates to the CDFs of luminous blue variables and WR stars. Additionally, we compare the type Ibn SNe to type II and type Ic CDFs from \cite{2012MNRAS.424.1372A}, presented in \cref{fig:anderson}. The measurements show that type Ibn associate with H$\alpha$ emission more strongly than either of the CCSN subtypes. We therefore conclude that massive stars, such as WRs, likely contribute significantly to the production of type Ibn SNe. While our results cannot reflect on the possibility of binaries as progenitor systems, they do signify the importance of very high mass stars in such systems.

\begin{figure}
    \centering
    \includegraphics[width=0.5\linewidth]{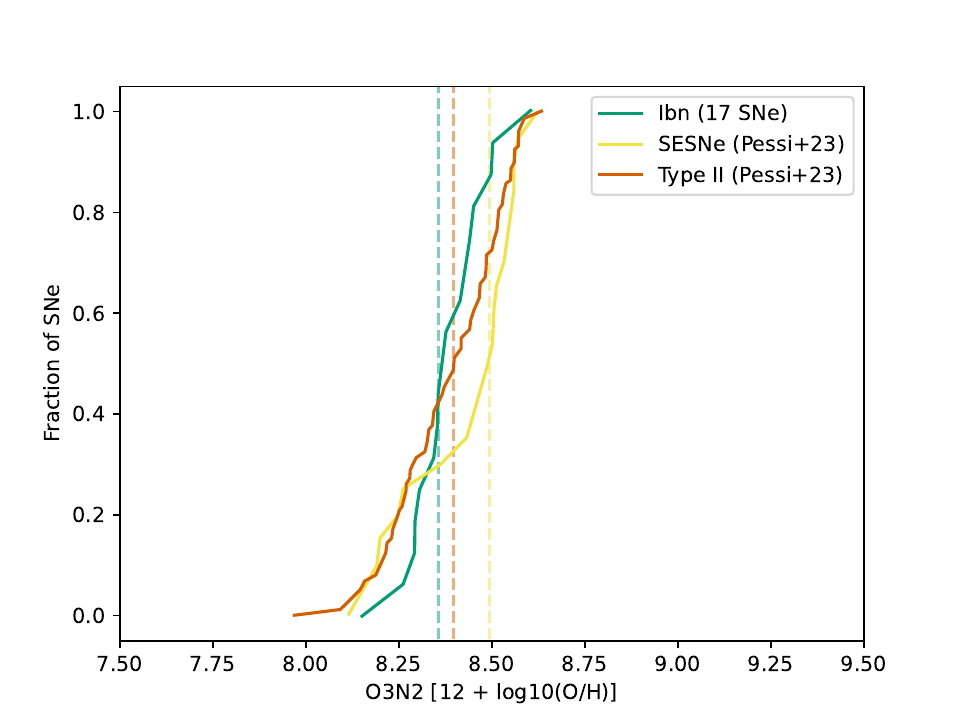}
    \caption{Type Ibn metallicity measurement in the N2 index compared to type II adapted from \cite{2023A&A...677A..28P}. Dashed vertical lines represent the median values of each SN type.}
    \label{fig:metals}
\end{figure}

We measured the metallicity of 17 SN locations in our sample, presented in \cref{fig:metals}. The metallicity of type Ibn SNe is comparable to the metallicity of type II SNe. We do not find any clear preference to high or low metallicity environments between type Ibn SNe or other SN types. This corroborates to earlier findings in \cite{2023A&A...677A..28P}, where they found no statistically significant difference between CCSN types in the N2 index.

For future work, we will attempt to increase our sample of metallicity measurements. By increasing the spectral extraction aperture we will be able to increase the signal-to-noise ratio of the SN environment at the expense of the size of the studied environment. This will also allow us to use fainter emission lines and use additional diagnostic tools, such as O3N2 or S2 as defined in \cite{2016Ap&SS.361...61D}. Additionally, statistical comparison of the type Ibn to the metallicities of other CCSNe is still needed.

\begin{thebibliography}{}
\bibitem[Anderson \& James(2008)]{2008MNRAS.390.1527A} Anderson, J.~P., \& James, P.~A.\ 2008, \textit{MNRAS}, 390, 1527.
\bibitem[Anderson, Habergham, James, \& Hamuy(2012)]{2012MNRAS.424.1372A} Anderson, J.~P., Habergham, S.~M., James, P.~A., \& Hamuy, M.\ 2012, \textit{MNRAS}, 424, 1372.
\bibitem[Dopita, Kewley, Sutherland, \& Nicholls(2016)]{2016Ap&SS.361...61D} Dopita, M.~A., Kewley, L.~J., Sutherland, R.~S., \& Nicholls, D.~C.\ 2016, \textit{Ap\&SS}, 361, 61.
\bibitem[Guillochon, Parrent, Kelley, \& Margutti(2017)]{2017ApJ...835...64G} Guillochon, J., Parrent, J., Kelley, L.~Z., \& Margutti, R.\ 2017, \textit{ApJ}, 835, 64.
\bibitem[Kangas et al.(2017)]{2017A&A...597A..92K} Kangas, T., Portinari, L., Mattila, S., et al.\ 2017, \textit{A\&A}, 597, A92.
\bibitem[Kennicutt(1998)]{1998ARA&A..36..189K} Kennicutt, R.~C.\ 1998, \textit{ARAA}, 36, 189.
\bibitem[Kilpatrick et al.(2021)]{2021MNRAS.504.2073K} Kilpatrick, C.~D., Drout, M.~R., Auchettl, K., et al.\ 2021, \textit{MNRAS}, 504, 2073.
\bibitem[Kuncarayakti et al.(2015)]{2015A&A...579A..95K} Kuncarayakti, H., Maeda, K., Bersten, M.~C., et al.\ 2015, \textit{A\&A}, 579, A95.
\bibitem[Marino et al.(2013)]{2013A&A...559A.114M} Marino, R.~A., Rosales-Ortega, F.~F., S{\'a}nchez, S.~F., et al.\ 2013, \textit{A\&A}, 559, A114.
\bibitem[Pessi et al.(2023)]{2023A&A...677A..28P} Pessi, T., Prieto, J.~L., Anderson, J.~P., et al.\ 2023, \textit{A\&A}, 677, A28.
\bibitem[West, Lauberts, Jorgensen, \& Schuster(1987)]{1987A&A...177L...1W} West, R.~M., Lauberts, A., Jorgensen, H.~E., \& Schuster, H.~E.\ 1987, \textit{A\&A}, 177, L1.
\end{thebibliography}
\end{document}